# Electrically Reconfigurable Arbitrary Splitting-Ratio Optical Splitter Based on Low-Loss Sb$_2$Se$_3$

**Yuru Li**[1,2,†], **Wanting Ou**[1,†], **Songyue Liu**[3,†], **Shunyu Yao**[1], **Ning Zhu**[4,8], **Qi Lu**[3], **Yuan Zhong**[3], **Lu Sun**[3,9], **Yan Li**[1,2,10], **Ying Li**[5], **Tao Zhang**[1], **Zhaohuan Ao**[1], **Zhaohui Li**[1,6], **Chao Lu**[7] **And Zhiyi Yu**[2]

[1]*Guangdong Provincial Key Laboratory of Optoelectronic Information Processing Chips and Systems, School of Electrical and Information Technology, Sun Yat-sen University, Guangzhou 510275, China*
[2]*School of Microelectronics Science and Technology, Sun Yat-sen University, Zhuhai 519000, China*
[3] *State Key Laboratory of Photonics and Communications, Department of Electronic Engineering, Shanghai Jiao Tong University, Shanghai, China*
[4]*School of Electronic Science and Engineering (School of Microelectronics), Center of Low Carbon and New Energy Materials, South China Normal University, Foshan 528225, China*
[5]*International Collaborative Laboratory of 2D Materials for Optoelectronic Science & Technology of Ministry of Education, Institute of Microscale Optoelectronics, Shenzhen University, Shenzhen 518060, China*
[6]*Southern Marine Science and Engineering Guangdong Laboratory (Zhuhai), Zhuhai 519000, China*
[7]*Photonics Research Institute, Department of Electronic and Information Engineering, The Hong Kong Polytechnic University, Hong Kong, SAR, China*
*†These authors contributed equally.*
[8]*Zhuning@scnu.edu.cn*
[9]*sunlu@sjtu.edu.cn*
[10]*liyan329@mail.sysu.edu.cn*

**Abstract:** Reconfigurable beam splitters capable of being arbitrarily programmed for the power splitting ratios are vital for the adaptive optical networks and photonic computing. Conventional mechanisms such as thermo-optic, free-carrier, or mechanical tuning are usually volatile and require continuous power, limiting their suitability for low-frequency and low power-consumption programmable operations. Here, we experimentally demonstrate an electrically reconfigurable beam splitter based on the low-loss phase-change material Sb$_2$Se$_3$, enabling multi-level and arbitrary splitting-ratio (SR) control. By locally triggering phase transitions in the coupling region with integrated micro-electrodes, we exploit the high refractive-index contrast between different phases and negligible absorption in the near-infrared wavelength of Sb$_2$Se$_3$ to precisely tune the coupling strength with non-volatile retention. 8-level of power splitting states is achieved within a compact footprint of ~14.5 μm in the experiments, with insertion loss is ~1 dB across 1515-1550 nm and near-zero static power. Combining the advantages of compactness, broad bandwidth, low loss, non-volatility, and multi-level control experimentally, this device provides a universal building block for scalable, energy-efficient reconfigurable photonic circuits, with great prospects in optical computing and intelligent communication systems.

## 1. Introduction

Integrated photonics, with its advantages of miniaturization, low power consumption, and large-scale integration, plays a central role in various applications such as the optical communication networks [1], data center interconnects [2], programmable photonic computing [3], and quantum information processing [4], etc [5,6]. Within this context, beam splitters with dynamically tunable splitting ratios may serve as the fundamental building blocks for adaptive optical networks [7], programmable photonic processors [8], and neuromorphic photonic computing [9]. They enable the adjustable and precise power distributions between multiple channels, not only supporting flexible routing and load balancing, but also underpinning the realization of optical logic and efficient resource scheduling [10,11].

Currently the tuning mechanisms of the photonic integrated circuits (PICs) [12] mainly rely on the thermo-optic effects [13], free-carrier dispersion [14], and micro-electro-mechanical (MEMS) [15]. While these approaches can provide high modulation speeds or large tuning depths, they are inherently volatile, meaning continuous power consumption which is not energy-efficient and also thermally unfriendly especially in the dense integration, thereby limiting their feasibility for applications that require low switching frequencies, persistent functionality, or long-term state retention without constant energy input [16-18].

Although thermo-optic and electro-optic splitters have been extensively studied at home and abroad, their implementations often suffer from large device size, limited bandwidth, and relatively narrow tuning ranges, making it difficult to realize the compact and broadband arbitrary ratio beam splitters [19-23]. These limitations underscore the need for alternative solutions, among which phase-change materials have recently emerged as a promising candidate [24,25].

Chalcogenide phase-change materials (PCMs) offer a promising route toward low-power reconfigurable devices [26]. Their distinctive properties include non-volatility, reversible switching between the amorphous and crystalline state, large refractive index contrast, and broad optical transparency [27-30]. PCM switching can occur at nanosecond-to-microsecond timescales with energy requirements ranging from picojoules to nanojoules, making them significantly more energy-efficient than continuously biased thermo-optic devices [31]. Although $Ge_2Sb_2Te_5$ (GST) has been widely demonstrated in various of integrated photonic devices such as switches [28,32-34], modulators [35-37], and tunable resonators [38-40], its severe absorption in the near-infrared (NIR) communication band introduces high insertion losses, restricting cascaded operation and large-scale deployment. By contrast, antimony selenide ($Sb_2Se_3$) has recently emerged as a low-loss PCM well-suited for the integrated photonics [41,42]. $Sb_2Se_3$ combines a near-zero extinction coefficient in the telecom band with a high refractive index contrast (~0.9 between amorphous and crystalline phases) [43,44], please refer to Supporting Information S1. Recent demonstrations of the $Sb_2Se_3$-based optical switches have exhibited prominent performances including broadband operation exceeding 50 nm, insertion loss as low as 1 dB, extinction ratios near 15 dB, and stable cycling performance beyond 1,000 switching events, highlighting its strong potential for practical reconfigurable circuits [45].

In this work, we exploit the favorable optical properties of $Sb_2Se_3$ to design and experimentally realize an electrically reconfigurable beam splitter with arbitrary splitting-ratio. The device is based on a silicon rib-waveguide directional coupler upon which a thin $Sb_2Se_3$ layer is precisely integrated. Its phase transition is controlled using the micro/nano-scale heating electrodes, allowing modulation of the effective indices of the coupled modes and enabling arbitrary control of the output splitting ratios. Leveraging the high index contrast and negligible absorption in the NIR region of $Sb_2Se_3$, the proposed device achieves low-loss, multi-level tunability within an ultra-compact footprint of only 14.5 μm. Experimental characterization shows an insertion loss of ~1 dB across 1515-1550 nm; using 20-μs electrical pulses to drive partial phase transitions in $Sb_2Se_3$, the device realizes eight programmable beam-splitting states. These results not only validate the feasibility of an arbitrary splitting-ratio tuning but also establish a versatile, compact, and energy-efficient platform for reconfigurable beam splitters in optical communication networks, data center interconnects, and programmable photonic computing.

## 2. RESULTS

### A. Design and Fabrication

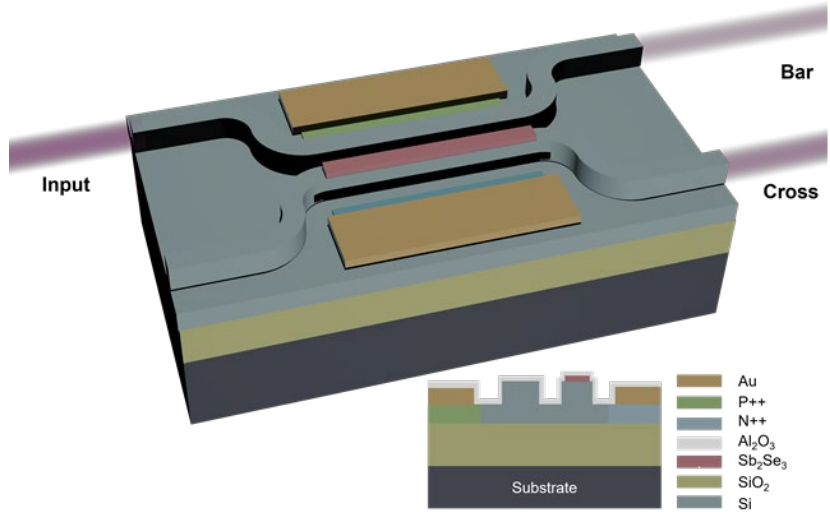

Fig. 1. Schematic diagram of the electrically reconfigurable beam splitter with arbitrary SR. The device is implemented on a silicon-on-insulator (SOI) platform and consists of two rib waveguides forming a directional coupler. A thin layer of $Sb_2Se_3$ is selectively deposited on the top of one waveguide, forming a hybrid waveguide structure that enables tunable coupling. Gold electrodes are integrated above the coupling region to serve as the localized heaters, while the underlying rib geometry supports PIN junction doping for efficient electrical control. The inset illustrates a cross-sectional view of the device, showing the entire layered structure.

The proposed device is implemented on a silicon-on-insulator (SOI) platform and employs a rib-waveguide directional coupler as the fundamental structure, as shown in Fig. 1. The rib geometry, defined by a partial-etch depth of 130 nm into the 220 nm silicon layer, offers two advantages: (i) it preserves low-loss guiding while maintaining strong optical confinement, and (ii) it facilitates the introduction of the PIN junction doping, enabling the integration of micro-heaters above the coupling region. Electrical characterization of the final PIN structure (see Supporting Information S2) confirms normal I-V behavior and proper functionality; although the series resistance is elevated due to fabrication tolerances, it does not compromise overall operation. To enable an active tunability, a thin film of $Sb_2Se_3$ is selectively deposited on the top of one of the rib waveguides, forming a hybrid waveguide structure. The $Sb_2Se_3$ thickness is set to 30 nm to balance the strong optical index modulation and fabrication practicality. To ensure precise alignment and reliable coverage during the lift-off process, the lateral width of the $Sb_2Se_3$ strip is designed to be 40 nm slightly narrower than the underlying silicon rib. The combination of silicon and $Sb_2Se_3$ provides a hybrid guiding medium whose effective index can be dynamically controlled by inducing phase transitions in the chalcogenide layer. Gold electrodes are integrated above the coupler and insulated by a dielectric cladding, providing localized resistive heating for phase transition control. The complete fabrication process is detailed in Supporting Information S3.

The design was first examined using eigenmode simulations performed in the MODE solver of Ansys Optics. Figs. 2 (a) shows the dependence of the effective index on waveguide width for pure silicon, amorphous $Sb_2Se_3$ (a-$Sb_2Se_3$) on Si, and crystalline $Sb_2Se_3$ (c-$Sb_2Se_3$) on Si, respectively. The results indicate that phase matching can be achieved when the silicon rib waveguide width is 500 nm and the hybrid waveguide width is 440 nm. This ensures that the coupling region supports efficient power transfer between the two arms when $Sb_2Se_3$ is in the amorphous state. The coupling gap was then optimized by analyzing the effective index difference $\Delta n_{eff}$ of the two lowest-order supermodes. As shown in Figs. 2 (b), $\Delta n_{eff}$ in the crystalline state is nearly twice of that in the amorphous state for the chosen gap. This chosen contrast is critical: in the amorphous state, when phase matching occurs, the optical power completely transfers to the Cross port at a coupling length $L_x = \lambda/(2\Delta n_{eff\_a})$ [46,47]. When Sb₂Se₃ transitions to the crystalline phase, the increased $\Delta n_{eff}$ detunes the coupling, the mismatched phase condition makes little light coupled into the Cross port. Besides, since the coupling length is reduced to nearly its one-half, this small part of leaky energy will be coupled back to the input waveguide at a length of $2L_x$, ensuring almost entire optical power transmission to the Bar port at the same device length (please refer to Supporting Information S4). In Fig. 2 (c), efficient coupling occurs in the amorphous state due to strong mode overlap, while in the crystalline state the field localizes within the hybrid waveguide, suppressing coupling and enabling port switching.

Finally, the device was evaluated for intermediate crystallization states by introducing a crystallization fraction m (please refer to Supporting Information S5) ranging from 0 (fully amorphous) to 1 (fully crystalline) [48,49]. As shown in Fig. 2 (d), the simulated transmission at 1550 nm exhibits a smooth evolution of the power SR: optical power gradually shifts from the Cross port to the Bar port as m increases, enabling multiple intermediate splitting states. This continuous transition demonstrates that the device is not limited to binary switching, but supports multi-level and arbitrary splitting-ratio control which is essential for programmable photonic circuits and reconfigurable optical interconnects.

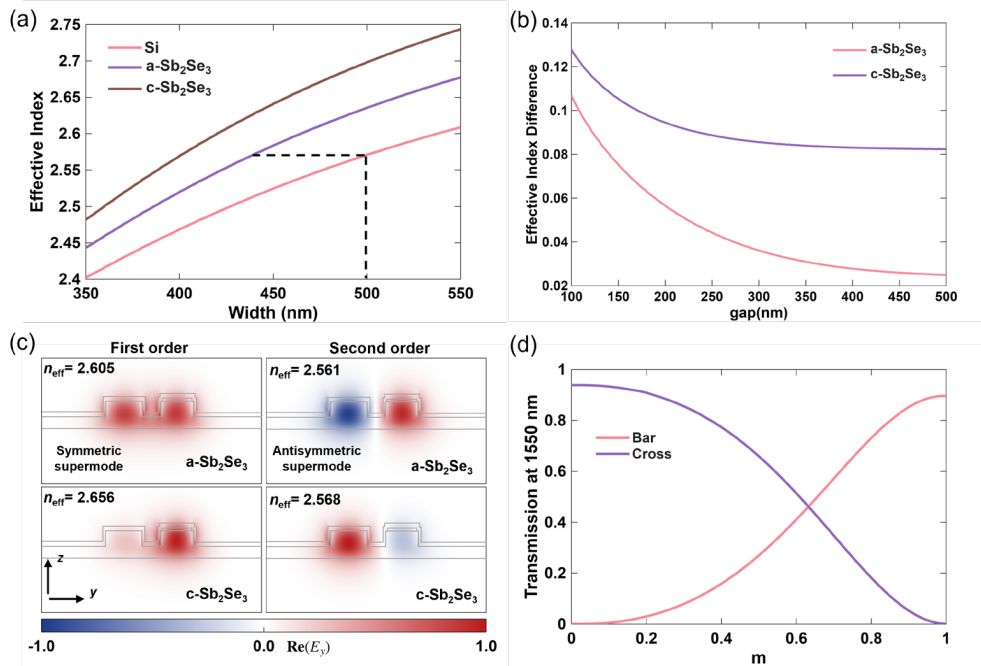

Fig. 2. Design methodology of the tunable splitting-ratio directional coupler. (a) Simulated effective indices of a pure silicon waveguide and an Sb₂Se₃-hybrid waveguide in both amorphous and crystalline states. The dashed line indicates the phase-matching condition, where a 500-nm-wide silicon waveguide corresponds to a 440-nm hybrid waveguide. (b) Calculated effective index differences of the two lowest-order supermodes as a function of the coupling gap for the structure with the amorphous and crystalline Sb₂Se₃ top layer, respectively. (c) Simulated mode profiles of the first two supermodes in the amorphous and crystalline states, illustrating the transition from symmetric/antisymmetric coupling to strong localization in the hybrid waveguide. (d) FDTD-optimized transmission characteristics at 1550 nm under different crystallization fractions of Sb₂Se₃, demonstrating continuous tunability of the splitting ratio.

We simulated the optical field evolution and transmission spectra of the proposed device under different crystallization fractions of Sb₂Se₃ across the wavelength range of 1500-1600 nm, with representative results shown in Fig. 3. The top row (Fig. 3 (a), Fig. 3 (c), Fig. 3(e)) presents the normalized electric field intensity distributions, while the bottom row (Fig. 3 (b), Fig. 3 (d), Fig. 3 (f)) shows the corresponding transmission spectra at the Bar and Cross ports. In the fully amorphous state (Fig. 3 (a), Fig.3 (b)), efficient phase matching is achieved between the two rib waveguides, and the optical power is almost completely transferred to the Cross port. The transmission spectrum remains flat across the 1500-1600 nm range, with the Cross port maintaining nearly unity transmission and the Bar port suppressed to below 0.05. This corresponds to an extinction ratio close to -30dB at 1550 nm, confirming the low-loss and high-extinction behavior of the amorphous state. When the $Sb_2Se_3$ layer undergoes partial crystallization $m = 0.63$ (Fig. 3 (c), Fig. 3 (d)), the effective index of the hybrid waveguide increases, partially detuning the phase-matching condition. As a result, the transmission reaches an approximately balanced state at 1550 nm, with ~50% of the optical power directed to each output, corresponding to a splitting ratio close to 1:1. In the fully crystalline state (Figs. 3 (e), Fig. 3 (f)), the effective index difference between the two supermodes nearly doubles as that of the amorphous state, resulting in a strong mismatch of the coupling condition. Consequently, the optical field is almost entirely confined within the hybrid waveguide, and the output is routed predominantly to the Bar port. The transmission spectrum confirms this behavior, with the Bar port exhibiting >0.95 transmission and the Cross port nearly extinguished across the entire wavelength range and also an extinction ratio of -30dB level at 1550 nm, demonstrating complete port switching reversal and thus validating the design principle. Taken the above results together, the device can continuously evolve from Cross-dominant output (amorphous state) to balanced splitting (partially crystallized) and finally to Bar-dominant output (crystalline state) by varying the crystallization fraction of $Sb_2Se_3$. This smooth transition confirms the ability of the proposed coupler to realize arbitrary and programmable splitting ratios with low insertion loss, establishing a powerful platform for reconfigurable integrated photonics.

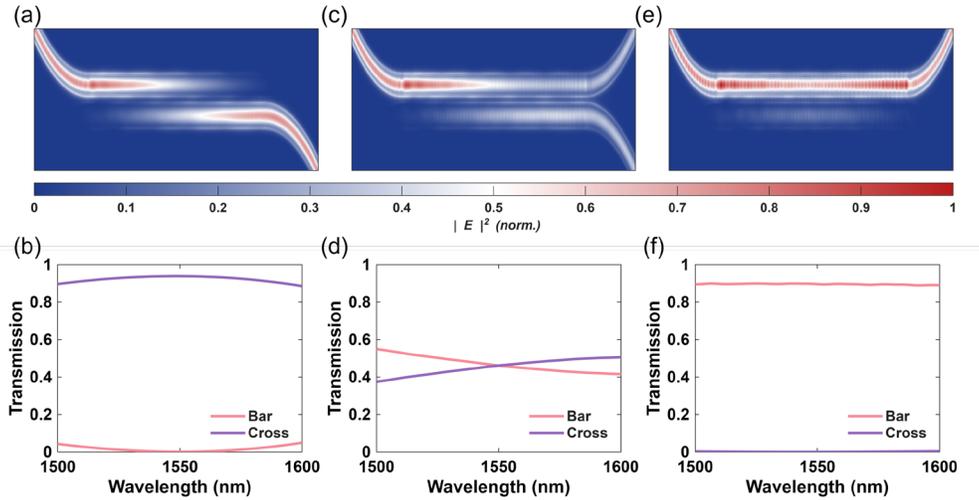

Fig. 3. Optical field distributions and transmission spectra of the device under different crystallization fractions of $Sb_2Se_3$. (a-b) Simulated electric field distribution and corresponding transmission spectrum in the amorphous state, showing efficient phase matching and nearly complete power transfer to the Cross port. (c-d) Device field distribution and transmission spectrum under partial crystallization, where the effective index detuning leads to balanced power splitting between the Cross and Bar ports. (e-f) Field distribution and transmission spectrum in the fully crystalline state, illustrating strong mode confinement in the hybrid waveguide and predominant power output from the Bar port.

### B. Experimental Results of the Optical Splitter

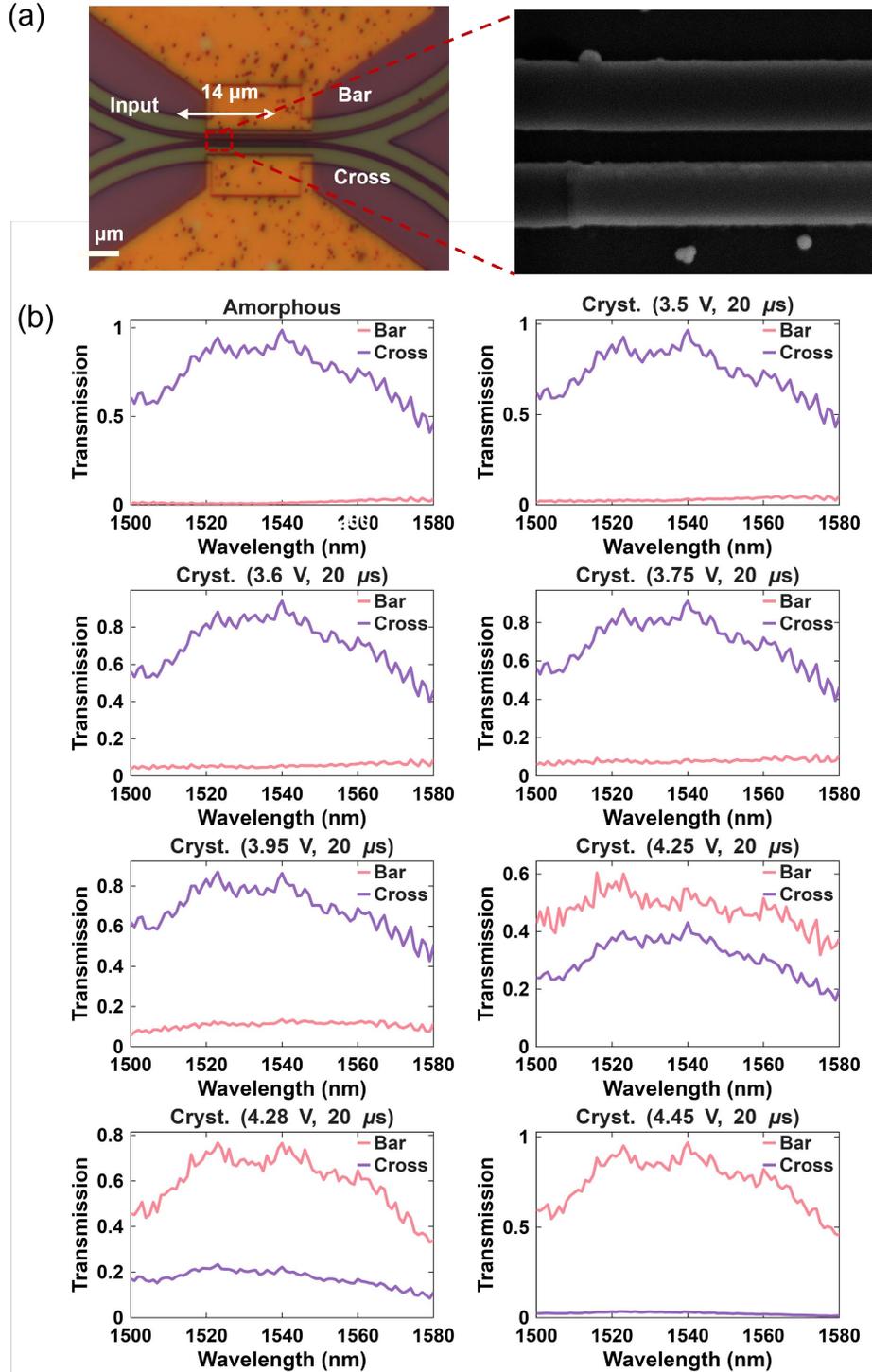

Fig. 4. Experimental characterization of the tunable beam splitter. (a) Optical microscope and SEM images of the fabricated device. The microscope image shows the overall device layout and compact footprint, with the input, Cross, and Bar ports clearly identified. The SEM image reveals the hybrid structure of the coupler, where the Sb$_2$Se$_3$ layer, deposited via a lift-off process,

uniformly covers the top surface of the silicon rib waveguide, ensuring precise alignment and thickness uniformity. (b) Measured transmission spectra at the Cross and Bar ports under different crystallization voltages. In the amorphous state, the output is dominated by the Cross port, with a power ratio of approximately 100:1 between the two ports. As the crystallization degree increases, the optical power gradually redistributes between the outputs. In the fully crystalline state, nearly all optical power is transferred to the Bar port, yielding a power ratio of approximately 1:100. To further verify the device's performance before and after crystallization, we conducted comparative measurements of the transmission spectra and SEM morphology for the amorphous and crystalline states; detailed results are provided in Supporting Information S7. These results confirm that the device maintains a compact footprint, low insertion loss, and non-volatile operation, while enabling continuous and multi-level tunability of the splitting ratio.

After completing the device design, we fabricated and experimentally characterized the tunable beam splitter, the test platform and measurement setup are detailed in Supporting Information S6. Fig. 4 (a) presents the optical microscope and SEM images of the fabricated device. The optical microscope image highlights the overall device footprint, with the input, Cross, and Bar ports clearly labeled, and the coupling region measured as only ~14.5 μm in length. The high-resolution SEM image further reveals the composite hybrid structure, where the $Sb_2Se_3$ layer deposited via a lift-off process is seen to uniformly cover the top surface of one Si rib waveguide. This deposition method ensures precise alignment and consistent thickness across the coupling region, providing reliable optical modulation performance.

We then measured the transmission spectra at both output ports under different crystallization states of $Sb_2Se_3$. In the amorphous state, the device exhibits low-loss operation, with the insertion loss at 1540 nm measured to be ~0.3 dB. Nearly all the optical power is directed to the Cross port, with a measured extinction ratio of approximately -20dB, confirming efficient phase-matched coupling and excellent extinction in the initial state. As the phase-change material undergoes partial crystallization, induced by applying 20 μs electrical pulses with gradually increasing amplitudes, the transmission spectra evolve smoothly (Fig. 4 (b)). With increasing voltage amplitude, the transmission at the Bar port continuously rises while the Cross port transmission decreases correspondingly, indicating controllable redistribution of optical power between the two output ports.

At intermediate voltages (e.g., 3.6-4.25 V), the spectra show different transmission levels at the Cross and Bar ports, corresponding to a varying splitting ratio from 1:0 to 1:1. This intermediate regime directly demonstrates the multi-level tunability of the device, enabling arbitrary splitting states beyond binary switching. Finally, when the applied voltage reaches 4.45 V, $Sb_2Se_3$ is driven into the fully crystalline state. In this condition, the insertion loss at 1515-1550 nm remains as low as ~1 dB, while the output switches almost entirely to the Bar port, with the extinction ratio between the two ports reaches approximately -20dB, in good agreement with the theoretical predictions.

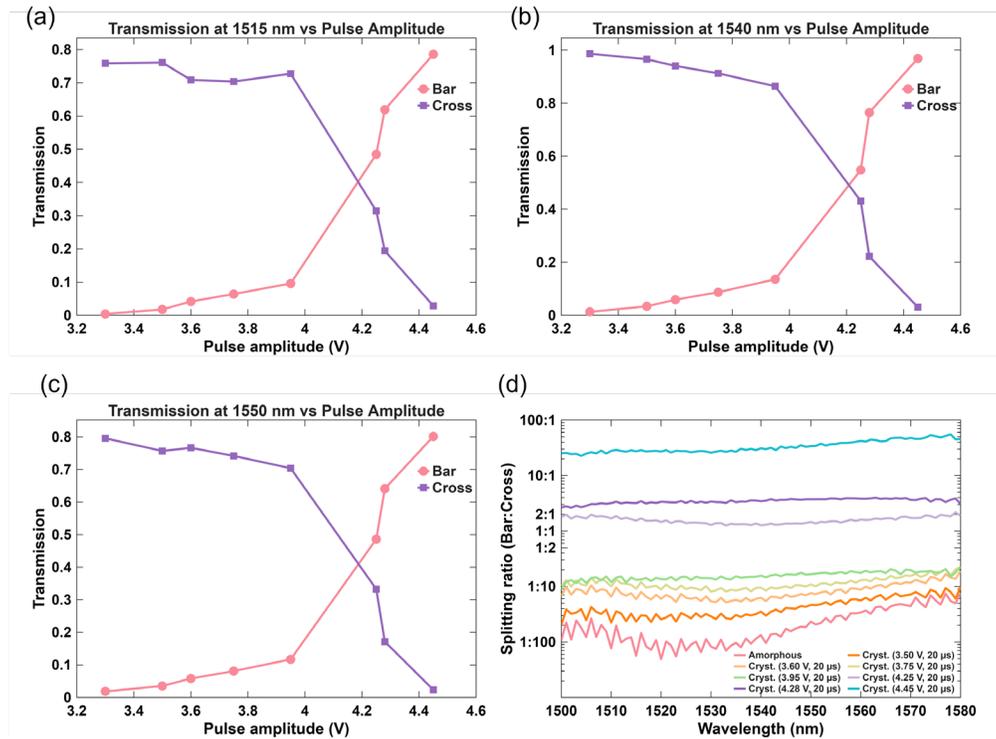

Fig. 5. Beam-splitting performance. (a-c) Bar and Cross transmissions versus pulse amplitude at 1515, 1540, and 1550 nm. With increasing voltage, power transfers smoothly from Cross to Bar, reaching a full port flip near ~4.45 V. (d) Broadband multi-level splitting ratio over 1500-1580 nm, plotted as Bar:Cross computed directly from the raw port powers (no extra normalization). The spectra are largely flat across the band; higher voltage shifts the ratio continuously from ~1:100 toward near 100:1. The legend denotes the amorphous state and several crystallized states (20 μs).

As shown in Fig. 5 (a-c), at three representative wavelengths—1515, 1540, and 1550 nm—the transmission of the two output ports exhibits a clear monotonic evolution with the drive voltage: as the voltage increases (i.e., the crystallization fraction grows), optical power smoothly transfers from the Cross port to the Bar port, with a complete port switch occurring near ~4.45 V. This behavior is consistent with the directional-coupling mechanism: when $Sb_2Se_3$ transforms from the amorphous to the crystalline state, its refractive index rises markedly, enlarging the effective-index difference of the two coupled-waveguide supermodes and thereby increasing phase mismatch and suppressing coupling. Consequently, for a fixed coupling length, the output gradually shifts from the Cross to the Bar port, enabling continuous tuning of the splitting ratio. Importantly, this monotonic, voltage-controlled behavior is reproducible across 1515-1550 nm, and the insertion loss of the multi-level states in this band is ~1 dB (please refer to Supporting Information S8), indicating strong bandwidth robustness. The small differences in splitting ratio among wavelengths primarily arise from the intrinsic dispersion of the directional coupler.

Fig. 5 (d) further presents multi-level splitting-ratio curves over a broad 1500-1580 nm range. The ratios are computed directly from the raw Bar and Cross powers as Bar:Cross (without any additional normalization). The spectral profiles for all states are generally flat and vary only weakly with wavelength, demonstrating a consistent power-distribution trend over an ~80 nm bandwidth. Meanwhile, as the voltage increases, the splitting ratio migrates continuously and with fine granularity between the two ports, spanning a wide range from ~1:100 to nearly 100:1. To further substantiate this conclusion, we list the splitting ratios at

1540 nm for all eight states; see Supporting Information S9. Although intrinsic coupler dispersion and minor fabrication-induced ripples introduce discernible shifts with wavelength, they do not compromise broadband, multi-level, and predictable control of the splitting ratio over 1500-1580 nm. These results validate the proposed design and highlight the hybrid $Sb_2Se_3$ directional coupler as a compact, low-loss, and reconfigurable on-chip beam splitter with strong potential for integrated photonics.

## 3. DISCUSSION

In this work, we propose and experimentally demonstrate an on-chip beam splitter with a tunable splitting ratio by tightly integrating the phase-change material (PCM) $Sb_2Se_3$ with a silicon directional-coupler (DC) architecture (a systematic comparison with recent beam-splitter reports is provided in Supplementary Information S10). $Sb_2Se_3$ combines a pronounced refractive-index contrast between its amorphous and crystalline phases, near-zero absorption across the telecommunications band, and inherent non-volatility, enabling ultra-low-energy steady-state retention. By precisely depositing $Sb_2Se_3$ on top of silicon waveguides, we form a hybrid DC in which appropriate voltages applied to the PCM layer modulate the effective indices of the supermodes in the coupling region, thereby accurately apportioning optical power between the two outputs and realizing arbitrary, programmable splitting ratios. The device occupies an ultra-compact ~14.5 μm footprint and maintains an insertion loss of ~1 dB over 1515-1550 nm across multiple crystallization states, owing to the excellent optical properties of $Sb_2Se_3$. Experimentally, by tuning the crystallization fraction of the PCM, we achieve eight discrete splitting states spanning approximately 100:1 to 1:100, including a near-balanced 1:1 state. To the best of our knowledge, this is the first on-chip beam splitter using a phase-change material to realize broadband, low loss, high-speed and arbitrary splitting-ratio tunability. Operated with microsecond/nanosecond-scale electrical pulses, the device delivers stable broadband, multi-level control while retaining a compact footprint, low loss, and non-volatile holding—highlighting the promise of the $Sb_2Se_3$-Si hybrid DC platform for programmable photonics and adaptive on-chip optical routing (see Supplementary Information for comparisons of electrically controlled PCM photonic devices).

Despite these encouraging results, the current implementation still has several limitations that require further improvement. Specifically, imperfections in the fabrication process restrict the overall reproducibility of device performance. Although the electrode design and placement are functional, they have not yet been fully optimized, which may lead to non-uniform heating or incomplete phase transitions. As the next step, further refinement of the fabrication process in the laboratory will be carried out, with a focus on improving electrode alignment accuracy and thermal management. In parallel, an optimized electrical driving schemes will be developed to achieve smoother and more continuous tuning of the crystallization fraction, thereby enabling stable and fully reconfigurable splitting ratio adjustment in future prototypes.

To look ahead $Sb_2Se_3$-based reconfigurable beam splitters hold strong potential for a wide range of applications, including intelligent optical networks, dynamic optical switching systems, and quantum communication, all of which demand flexible and accurate control of optical power distribution together with an excellent balance of compact footprint, broad bandwidth, low insertion loss, and non-volatility. While this work demonstrates the feasibility of the multi-level programmable splitting, practical deployment in large-scale photonic integrated circuits still requires enhanced precision and reproducibility of splitting states. Future research efforts should focus on achieving more accurate and repeatable electrical control of the crystalline fraction of PCM for controllable SR. Improvements in electrode design, thermal management, and material engineering are expected to enable more stable and fine-grained electrical tuning of splitting ratios. In summary, this study paves the way to the next-generation reconfigurable components in various applications such as large-scale optical interconnects and programmable optical computing by offering a versatile and energy-efficient building block for electrically-adjustable integrated photonics.

**Funding.** National Key R&D Program of China (No. 2023YFB2905500); Southern Marine Science and Engineering Guangdong Laboratory (Zhuhai) (No. SML2024SP006); National Natural Science Foundation of China (No. 62275162);

**Disclosures.** The authors declare no conflicts of interest.

**Data availability.** Data underlying the results presented in this paper are not publicly available at this time but may be obtained from the authors upon reasonable request.

**Supplemental document.** See Supplement 1 for supporting content.